\documentstyle[prb,aps,twocolumn,epsfig]{revtex}
\begin{document}
\newcommand{\bra}[1]{\langle #1 |}
\newcommand{\ket}[1]{| #1 \rangle}
\draft

\title{Marshall's Sign Rule and DMRG Acceleration}

\author{Ulrich Schollw\"{o}ck}
\address{Sektion Physik, Ludwig-Maximilians-Univ.\ M\"{u}nchen,
Theresienstr.\ 37, 80333 Munich, Germany}

\date{April 20, 1998}

\maketitle
\begin{abstract}
In applications of White's Density Matrix Renormalization Group (DMRG) 
algorithm,
computation time is dominated by the diagonalisation of large sparse 
Hamiltonians by iterative diagonalisation algorithms, whose convergence 
can be decisively accelerated by the usage of good start vectors.
In this paper I show how, using the Marshall sign rule,
in a wide class of antiferromagnetic models the number of diagonalization
iterations can be reduced below 10, sometimes down to 2, accelerating
the DMRG by an order of magnitude. This acceleration, applicable during
the growth of long chains, complements the acceleration procedure proposed
by White. To illustrate the feasibility of the approach, I show how it 
performs if applied to the calculation of the Haldane gap for
$S=2$.  
\end{abstract}

\narrowtext
\section{Introduction}
In recent years, the density matrix renormalization group algorithm proposed
by White\cite{White 92b} in 1992, 
has become together with exact diagonalisation and quantum
Monte Carlo methods the algorithm of choice in low-dimensional quantum
models, including magnetic, fermionic and bosonic systems. 

White has proposed two different DMRG algorithms,
the so-called finite-size and infinite-size algorithm. The 
infinite-size algorithm precedes the application of the finite-size algorithm:
a chain is grown symmetrically to its final size, whereas the finite-size
algorithm treats the fully-sized system splitting it asymmetrically, which
increases the precision of the results. Sometimes, it is however advantageous
to stick to the infinite-size algorithm and increase its precision by
increasing the number of block states $M$.
For example, when the Hamiltonian is invariant under reflection
and parity thus a good quantum number,
it is of great advantage to retain this quantum number, for
easier classification of states and for thinning out the Hilbert space by
splitting it into more sectors invariant under the operations of the
Hamiltonian. This gives access to more states 
and speeds up the algorithm. 

In either case, DMRG precision is dominated by the number of block states
$M$, and time consumption scales as $M^{3}$. The most time-consuming part of
the algorithm is the determination of low lying eigenstates of the approximate
Hamiltonians that are formed during the iterative application of the
decimation process of the DMRG. To do this, iterative diagonalisation
algorithms such as Lanczos\cite{Lanczos 50,Paige 72} are used. They can
be accelerated if a good initial guess for the targeted state is available,
as the number of (Lanczos) iterations drops.

During the application of the finite size algorithm, a previous
ground state for full-length system under consideration as well
as the (incomplete) DMRG basis transformations are available. 
White has used this information to make a very good guess for
the initial state of the diagonalization algorithm\cite{White 96b},
basically carrying out basis transformations on the previous ground state. 
During the application of the infinite size algorithm, previous
ground state(s) were all obtained on shorter chains, such that a basis
transformation is not feasible. What I want to show in this paper is
how in the infinite size algorithm the old information can be used to
also make a good initial guess.

What the DMRG essentially does is to find a fixpoint in density matrix
space, i.e.\, for large systems, essentially the same (incomplete)
transformation from a basis with $MN$ states ($M$ block states and $N$
spin states) to a new (decimated) basis with $M$ 
states is carried out\cite{Ostlund 95}.
A first guess would be that
the target state converges to a fixpoint in Hilbert space also, implying
that simply using the target state found in the last iteration as the
initial state vector should be a good guess. 

This is not so: the overlap between the initial guess and
the result is typically far from 1, and $\langle v_1 | {\cal H} | v_1 \rangle$,
the energy expectation value of the initial guess is much larger than the
true target state energy. One finds that for longer
chains, the absolute values of wave function coefficients (if labeled
suitably, see below) converge fast to
fixed values, while the signs vary randomly. If one could predict the sign
changes, the old wave function would indeed provide an excellent starting
point. The randomness in signs has two origins. First, real eigenvectors are
only determined up to a global sign, which is attributed unpredictably by the
density matrix diagonalization algorithm. Second, there are deterministic sign
changes with chain length. We will see how the latter can be used to fix the
global signs of eigenvectors such that almost all signs of the new wave
function are correctly reproduced in the old wave function, which then can
serve as excellent prediction for the new wave function.  

\section{Prediction Mechanism}
Let us consider, to simplify the description, an isotropic Heisenberg 
antiferromagnetic spin chain with integer spin length $S$. Let us 
assume that we have reached a certain block length $L$, such that the
total chain under consideration has length $2L+2$. Let us call the
block states of the block of length $L$ $\ket{m'_{L}}$, those of the
block of length $L+1$ $\ket{m_{L+1}}$, with total magnetisations
$S^{z}_{m'_{L}}$ and $S^{z}_{m_{L+1}}$. If we call $\sigma$ the spin state
on site $L+1$, and the magnetisation $S^{z}_{\sigma}$, the DMRG 
decimation procedure yields
\begin{equation}
\langle m'_{L} \sigma | m_{L+1} \rangle \neq 0 
\end{equation}
with 
\begin{equation}
S^{z}_{m'_{L}} + S^{z}_{\sigma} = S^{z}_{m_{L+1}}
\end{equation}
as necessary condition. Both the states of the block
with length $L$ and the block with length $L+1$ are now {\em ordered and 
numbered in the 
respective
magnetisation sectors according to their importance, i.e.\ the associated
eigenvalue of the density matrix when they were formed.} I assume now that
the respective magnetisation sectors for the blocks of length $L$ and $L+1$ 
have equal numbers of states. This is normally the
case as soon as chains exceed certain, rather small lengths; if not, typically
one or two states of extremely small weigth are redistributed. One can easily
modify the proposed procedure to take this into account, which for reasons
of clarity I will not discuss; all results shown below were obtained by
switching off the prediction mechanism in these rare circumstances.
Otherwise the condition could be
ensured by adding or dropping a state which has very little weight, at minor
loss of precision.

{\em Fundamental rule.}
If this assumption holds, the global sign of the eigenvector of the
density matrix giving $\ket{m_{L+1}}$ is chosen such that
\begin{equation}
(-1)^{S^{z}_{m_{L+1}}}\langle m_{L} \sigma | m_{L+1} \rangle > 0 
\label{eq:fund}
\end{equation}
for the case where $m_{L}\equiv m_{L+1}$, meaning that the states have
{\em equal label number} (I will call them ``equivalent''). 

\begin{figure}
\centering\epsfig{file=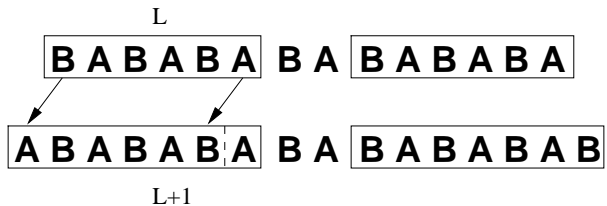,scale=0.8}
\vspace{0.3truecm}
\caption{DMRG chains for different block lengths $L$ and $L+1$ with sublattice
  structure. Sublattices are labeled such that central spins remain on the
  same sublattice. Blocks are represented by rectangles. Note the shift in
  sublattice position of block spins during system growth.}
\label{fig:DMRG}
\end{figure}

How can this rule be motivated? Let us consider Marshall's sign 
rule\cite{Marshall 55}. In its
original form it states that for a bipartite connected (i.e.\ that only
sites on different sublattices $A$ and $B$ interact and that each site
can be linked to each other site by a chain of bonds) isotropic 
antiferromagnet the lowest energy state $\ket{\psi^{M}}$ 
in a sector with total magnetisation $M$ can be written in the Ising basis
\begin{equation}
\ket{\psi^{M}} = \sum_{n} c_{n} (-1)^{\sum_{i\in B} -S + S^{z}_{i}} \ket{n},
\end{equation} 
such that $c_{n}>0$. The $\ket{n}$ are the Ising basis states, the sum
in the exponent effectively yields the total magnetisation of sublattice
$B$. Marshall's sign rule applies to the two most interesting states for the
DMRG, the ground and first excited states. Let us compare the two chains
arising in the DMRG procedure shown in Figure 1, for block length $L$ and
block length $L+1$. If we assign the same
sublattice positions to the center spins, the positions of the spins in the
$L$-block shift their position from one sublattice to the other. Let us now
take the two (say, ground state) wave functions
\begin{eqnarray}
& & \ket{\psi_{L}} = \sum \psi_{m^{l}_{L}\sigma^{l}\sigma^{r}m^{r}_{L}}
\ket{m^{l}_{L}\sigma^{l}\sigma^{r}m^{r}_{L}} \\
& & \ket{\psi_{L+1}} = \sum \psi_{m^{l}_{L+1}\sigma^{l}\sigma^{r}m^{r}_{L+1}}
\ket{m^{l}_{L+1}\sigma^{l}\sigma^{r}m^{r}_{L+1}} 
\end{eqnarray}
If we make now the assumption that the DMRG is close enough to its fix
point that we may identify both systems, the wave function coefficients
will be (almost) identical {\em but for the sign}, as sublattice positions
of spins have changed. This is confirmed by numerical evidence, and motivates
calling states equivalent; while physically distinct, they make (almost) the
same contribution to (physically also distinct) wave functions as also
supported by the arguments of Ref.\ \onlinecite{Ostlund 95}. 
It
remains to determine the sign. If we consider two equivalent states 
$\ket{m_{L}}$
and $\ket{m_{L+1}}$, the matrix element
$\langle m_{L} \sigma | m_{L+1} \rangle$ 
is non-zero only if $S^{z}_{\sigma}=0$. 
Expanding $\ket{m_{L}}$ in the Ising basis, the signs of the wave function
coefficients will, up to a global sign, be given by
$(-1)^{[S^{z}]^{B}_{n}}$, where $[S^{z}]^{B}_{n}$ is
the sum over the magnetisation on sublattice $B$ of Ising state $n$.
If we identify now $\ket{m_{L}}$ in $\ket{\psi_{L}}$ with $\ket{m_{L+1}}$
in $\ket{\psi_{L+1}}$, 
one
sees that the added spin (sitting on sublattice $A$) makes no sign 
contribution, while the spins of the $L$-block have changed sublattice,
such that the sign will be $(-1)^{[S^{z}]^{A}_{n}}$. The
relative sign change will be
\begin{equation}
(-1)^{[S^{z}]^{A}_{n}-[S^{z}]^{B}_{n}} =
(-1)^{[S^{z}]^{A}_{n}+[S^{z}]^{B}_{n}} =
(-1)^{S^{z}_{m_{L}}} = (-1)^{S^{z}_{m_{L+1}}}.
\end{equation} 
The last expression is independent of the underlying Ising states and leads
immediately to rule (\ref{eq:fund}).

{\em Majority rule.} Retracing the above argument, its central weakness is
the assumption that block states can be identified for different lengths.
Calculating the overlap between predicted wave functions and calculated
wave functions, it is found to have increased to typically well above $0.9$,
but this does not yet lead to a large increase in performance of iterative
diagonalizations. Thus, rule (\ref{eq:fund}) 
seems to catch an important point, but
is also oversimplifying. Block identification implied that the added
spin had zero magnetisation, which is physically not true; there will be other
magnetisations contributing, in particular in excited states. Rule
(\ref{eq:fund}) 
may
therefore not apply. 

To go beyond, magnetisations $S^{z}_{\sigma}\neq 0$ have to be considered
also. It seems that there is no strict rule to do this -- also because the overlaps
between states may be so small that the sign is up to numerical arbitrariness
-- but we have found the following procedure to work extremely well.

The expression
\begin{equation}
(m_{L+1},m'_{L+1}) := \mbox{sign\ }(-1)^{S^{z}_{\sigma}} \langle m'_{L}\sigma | m_{L+1}\rangle \langle
m_{L} (-\sigma) | m'_{L+1} \rangle ,
\end{equation}
where $\ket{m_{L}}$ and $\ket{m_{L+1}}$, and $\ket{m'_{L}}$ and 
$\ket{m'_{L+1}}$ are equivalent
states respectively, is evaluated. If it is negative, the sign is 
considered ``wrong''. One finds, that almost all states $\ket{m_{L+1}}$ have 
all
signs $(m_{L+1},m'_{L+1})$ right, 
and that some states (typically less than 10 out of
several hundred) have almost all signs wrong. The wrong signs of the
``almost correct'' states $\ket{m_{L+1}}$ 
are typically those $(m_{L+1},m'_{L+1})$, where $\ket{m'_{L+1}}$ is an
``almost wrong'' state. By globally flipping the sign of the few
``almost wrong'' states, nearly all signs can be made right; the majority
character of the rule makes it extremely stable against numerical
imprecisions. The fact that only few flips are necessary justifies the
assumptions made for (\ref{eq:fund}):
As we will show, the majority rule added to (\ref{eq:fund})
ensures that the old wave function has an almost perfect overlap
with the desired new wave function.

One of several ways to
see how this rule goes beyond (\ref{eq:fund}) is to imagine that we are
adding two sites to a block, assuming that $m_{L}$ and $m_{L+2}$ can be
identified, with spin $-\sigma$ on the first and spin $\sigma$ on the second
site. There is no shift in sublattice positions, one of the new spins
makes a sign contribution $(-1)^{S^{z}_{\sigma}}$. Therefore,
\begin{equation}
\mbox{sign\ } [(-1)^{S^{z}_{\sigma}}\langle m_{L}(-\sigma)\sigma | m_{L+2}\rangle] = 1 ,
\end{equation}
for identified states. Inserting 
$\sum_{m'_{L+1}}| m'_{L+1}\rangle \langle m'_{L+1}|$, we find
\[
\mbox{sign\ } [(-1)^{S^{z}_{\sigma}}\sum_{m' }\langle m_{L}(-\sigma)|m'_{L+1}\rangle
\langle m'_{L+1} \sigma | m_{L+2}\rangle] = 1 .
\]
Now we don't make a statement about 
$\langle m_{L}(-\sigma)|m'_{L+1}\rangle$ (this would not go beyond
(\ref{eq:fund})),
but about a product. Identifying the decimations $L\rightarrow L+1$ and 
$L+1 \rightarrow L+2$
in the above for large $L$\cite{Ostlund 95}, 
we obtain the majority rule at least for the dominant overlaps.

{\em Extensions.} So far, we have only considered isotropic chains. Retracing
the derivation of the Marshall sign rule, it becomes evident that
(antiferromagnetic) anisotropies and dimerized interactions are compatible
with the sign rule; we found the prediction to work\cite{Aschauer 98}. 
In other cases, such as frustration, where the Marshall
sign rule does not apply, Richter, Ivanov and Retzlaff have shown that for
small frustration the weight of the states whose signs violate the sign rule
is zero or very small\cite{Richter 94}. 
In such cases, the method will still lead to an
efficiency gain, which gradually disappears with increasing frustration.
This has been confirmed numerically. We
have not considered half-integer spins, as state parity alternates during
chain growth; this implies that states that are two DMRG steps apart will be
related, making the method more complicated. However, the sign rule still
holds, and it should be possible to adapt the prescription to that situation.

The whole procedure can be reformulated to leave the transformation matrix
invariant, and use it and Marshall's sign rule to change the ground state wave 
function. In any case, no new
information is introduced: all procedures that just use the results of the
last decimation inherently assume that one is already at the DMRG fixpoint, 
where (but for the sign) wave function coefficients do not change. To
go beyond, one has to evaluate this assumption, e.g.\ by studying whether the
last two incomplete transformations $m_{L-1} \rightarrow m_L$ and 
$m_L \rightarrow m_{L+1}$ are (loosely speaking) identical, 
leading eventually to expressions similar in spirit to Ref.\ 
\onlinecite{White 96b}.
To be able to compare meaningfully two independently computed 
transformations, with
random signs independently attributed by the diagonalization algorithms, a
sign-fixing procedure as presented in this paper has to be invoked in any
case, whereas in the finite-size algorithm the sign effects are already
incorporated in the wave function. 
Results are not presented here, as the further
improvement in convergence we
obtained was, though systematic, rather disappointing.
   
\section{Performance}
\begin{figure}
\centering\epsfig{file=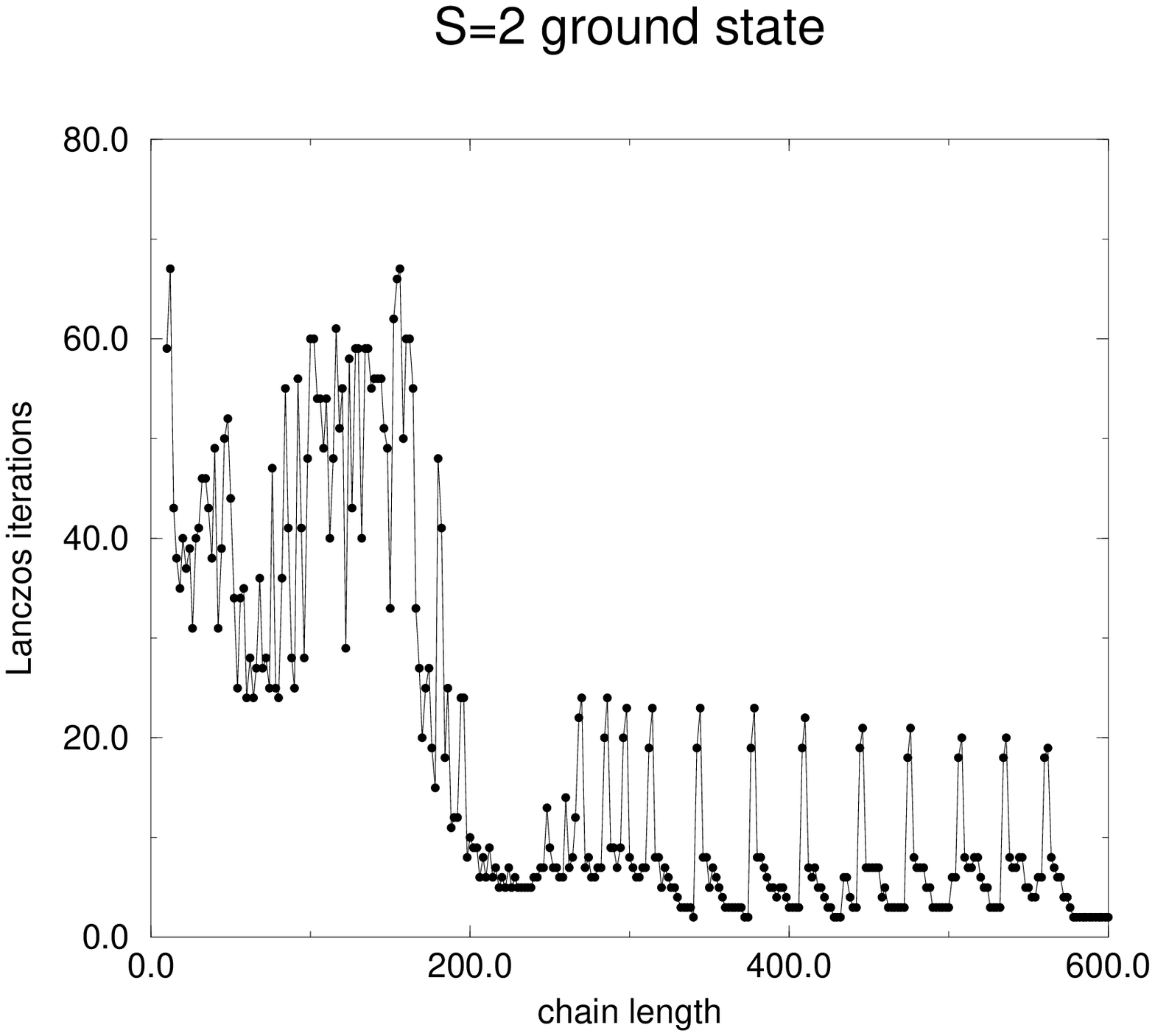,scale=0.5}
\vspace{0.3truecm}
\caption{Number of Lanczos iterations needed for the ground state of
a $S=2$ chain vs.\  length. Note that the minimum number of Lanczos
iterations to establish convergence is 2.}
\label{fig:s2gslan}
\end{figure}

The above two rules can be implemented very easily (30 or so lines at most),
and consume almost no memory and computation time. The gain is however
striking. To illustrate that the above procedure is useful, we calculated the
$S=2$ Haldane gap using chains with total length $L_{tot}=600$, 
while keeping $M=400$ block states, employing magnetisation and parity as
good quantum numbers. The chain is long enough, that the expected $L^{-2}$
convergence of the finite length gap can be observed, and a gap
$\Delta=0.0907(2)$ extrapolated (cf.\ Refs.\ \onlinecite{Schollwoeck 95}). 
The number of Lanczos iterations without
prediction varies between $\sim 60$ for the ground state and $\sim 70$ for the
first excitation.
As can be seen in Fig.\ \ref{fig:s2gslan}, the prediction makes the number of Lanczos iterations
drop 
dramatically, once the chain exceeds a certain length. For $L>200$, more
than 30 percent of all Lanczos
runs finish after 4 or less iterations, where it has to be kept in mind
that 2 is the minimum number of iterations to establish that a result has
already converged. Less than 15 percent take 10 or more iterations, and
the average is 7 iterations only. 

\begin{figure}
\centering\epsfig{file=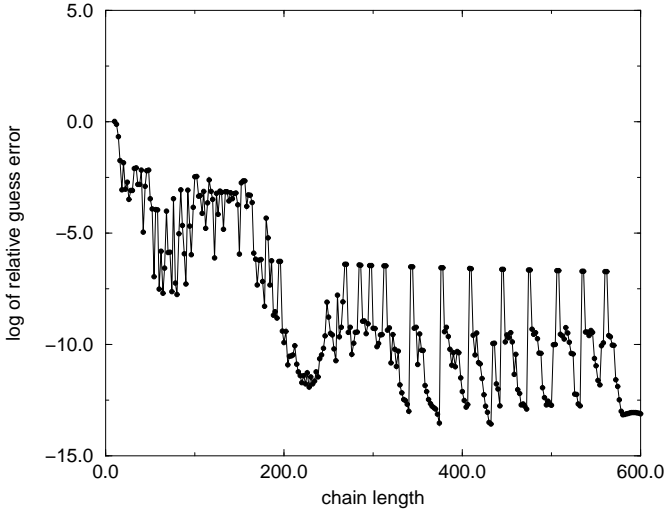,scale=0.5}
\vspace{0.3truecm}
\caption{Decadic logarithm of the relative guess error
  $(E_{final}-E_{guess})/E_{final}$ for the ground state
of a $S=2$ chain vs.\ length. }
\label{fig:s2gserr}
\end{figure}

The intermittent peaks are due to rearrangements of the
block states which reflect the growing chain length. In these iterations,
prediction works only partially (the same phenomenon can be observed in strong
form below $L=200$). The number of Lanczos iterations is
closely connected to the guess error (Fig.\ \ref{fig:s2gserr}), 
the difference between the energy
expectation value for the reused old ground state and the converged Lanczos
result. For $L>200$, the guess error relative to the ground state energy 
is always better than $10^{-6}$, mostly at or below
$10^{-10}$. This shows that the prediction rules are very powerful
indeed. Remaining errors also come from the fact 
that the assumption that the absolute
weight of states does not change 
with chain length is of course only approximate.

\begin{figure}
\centering\epsfig{file=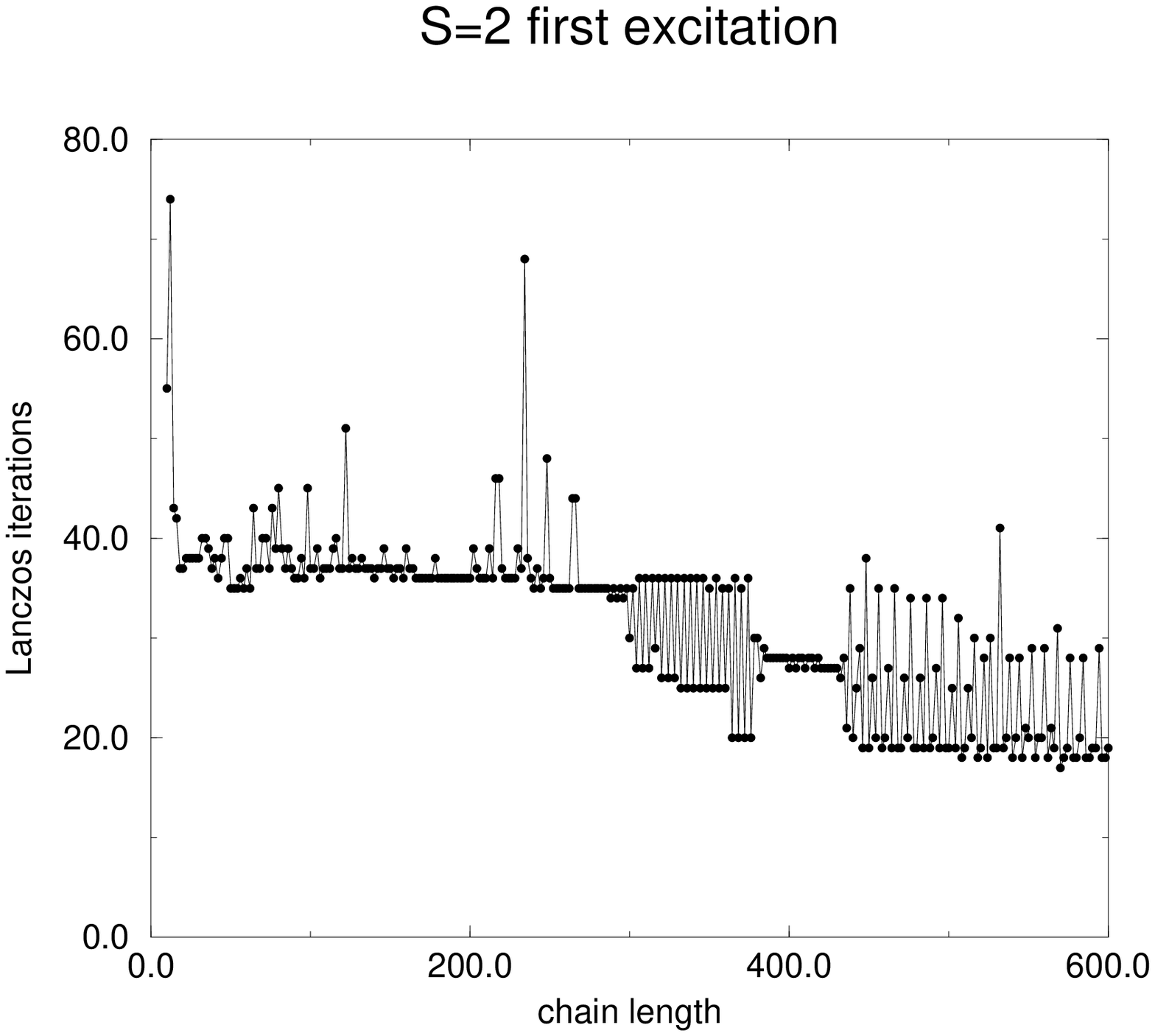,scale=0.5}
\vspace{0.3truecm}
\caption{Number of Lanczos iterations needed for the first excitation of
a $S=2$ chain vs.\ length. }
\label{fig:s2exlan}
\end{figure}

\begin{figure}
\centering\epsfig{file=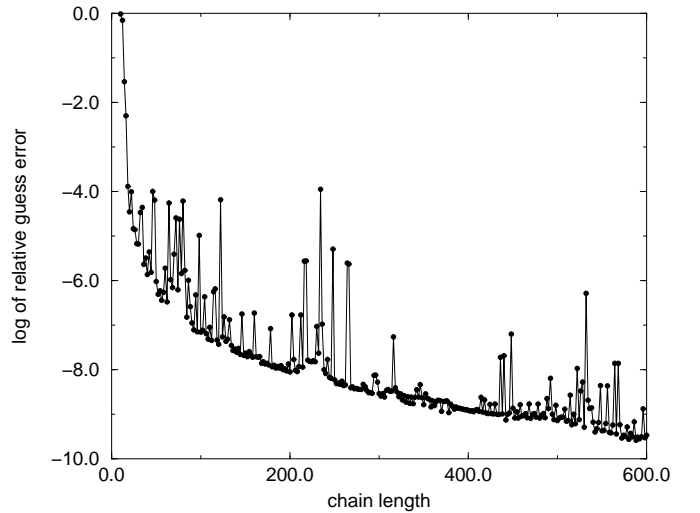,scale=0.5}
\vspace{0.3truecm}
\caption{Decadic logarithm of the relative guess error
  $(E_{final}-E_{guess})/E_{final}$ for the first excitation
of a $S=2$ chain vs.\ length. }
\label{fig:s2exerr}
\end{figure}

Results for the first excitation are not quite as good, as was to be expected,
but still the number of Lanczos iterations drops to about 20, with some peaks
up to 40. This is still a saving up to a factor 3. The relative error goes
down to $10^{-9}$, which shows that the method is also efficient
here. 

Let us close with the remark that on an alpha workstation, these calculations
could all be done overnight and that $S=2$ is not an ``easy'' case for
demonstration purposes: the very long correlation length is damaging for the
underlying assumptions of our procedure.

\section{Conclusion}
We have shown how using information about the nature of antiferromagnetic wave
function provided by the Marshall sign rule can be used to strongly reduce the
number of diagonalization steps in the infinite system growth phase of the
DMRG, allowing the growth of very long high-precision chains in reasonable
time. While the proposed procedure is not as completely versatile as White's
finite chain procedure\cite{White 96b}, 
it covers many important scenarios and can be used as
complementary to his procedure.

{\em Acknowledgements.} I thank J. Richter for drawing my attention to his
continuity argument for the Marshall rule in frustrated chains.

\end{document}